\documentclass[12pt]{article}
\usepackage[dvips]{graphicx}
\usepackage{epsf}
\usepackage[T2A]{fontenc}
\usepackage[english]{babel}

\begin{document}
\title{Conversion of Dark matter axions to photons in magnetospheres of neutron stars}
\author{M.S. Pshirkov\footnote{E-mail: pshirkov@prao.ru}, S.B. Popov}
\date{\emph{PRAO of Astro Space Centre of the P.N.Lebedev Physics Institute\\
Sternberg Astronomical Institute}}

 \maketitle

\textbf{Abstract.} We propose a new method to detect observational
appearance of Dark Matter axions. The method utilizes observations
of neutron stars (NSs) in radio. It is based on the conversion of
axions to photons in strong magnetic fields of NSs (Primakoff
effect). Whether the conversion takes place, the radio spectrum of
the object would have a very distinctive feature -- a narrow spike
at a frequency corresponding to the rest mass of the axion. For
example, if the coupling constant of the photon-axion interaction
is $M=10^{10}$ GeV, the density of Dark Matter axions is
$\rho=10^{-24} \, {\rm g cm^{-3}}$, and the axion mass is  $5 \, {\rm \mu
eV}$, then a flux from a strongly magnetized
($10^{14}$~G) NS at the distance  300 pc from the Sun
is expected to be about few tenths of mJy
at the frequency $\approx 1200$ MHz in the bandwidth $\approx 3$
MHz. Close-by X-ray dim isolated neutron stars are proposed as
good candidates to look for such radio emission.

\section{Introduction}
Most of the matter in the Universe is "dark", i.e. it can not be
observed directly  by astronomical observations, as the particles
that form it are not baryons \cite{peebles}, \cite{kamionkowski}.
A lot of types of new particles were suggested by the theorists to
explain the Dark Matter (DM) problem. One of the best candidates
with a strong theoretical background is axion -- a light neutral
pseudoscalar particle, which appears in spontaneous breaking of
Peccei-Quinn symmetry \cite{pq} as a solution of the strong CP
problem.

In fact, axions are not really dark. In external electromagnetic
field they can couple to  virtual photons and produce  real
photons (so-called Primakoff effect \cite{prim})

\begin{equation}
\label{primakoff}
 a+\gamma_{\rm virt} \rightarrow\gamma
\end{equation}

Several experiments that utilize the Primakoff effect are underway
now. One direction of such studies is to look for an effect
related to conversion of high energy axions from interiors of  the
Sun in strong magnetic fields in a laboratory \cite{CAST},
\cite{Tokyo}. Also solar axions can be searched in X-ray
observations because of  conversion of these particles in the
Earth magnetic field \cite{ziotas},\cite{davoud}. All these
experiments are searching for   "hot" and "young" axions, not
"cold" cosmological particles originated in the Early Universe.
Here we propose an astronomical method to detect emission due to
conversion of cosmological axions. Direct experimental searches
for such axions are in progress now in laboratories, too
\cite{sik2}.

The  strongest magnetic fields known can be found in the
surroundings of neutron stars (NS), so under certain conditions
axions that constitute cosmological DM can experience the
Primakoff effect in magnetospheres of NS. Possible use of such
super-strong magnetic fields for searching light pseudo-scalar
bosons in processes of photon conversion was previously discussed
in several papers \cite{gnedin1,gnedin2}. In what follows, a
possibility of detection of photons from such conversion  is
studied.

\section{Theoretical model}

The axion-photon  coupling is given by the following term in a
lagrangian:
\begin{equation}
\label{lagranzhian}
 L_{\gamma\phi}=-\frac{1}{4M}F^{\mu\nu}\tilde{F}_{\mu\nu}\phi,
\end{equation}
where $\phi$ -- the axion field strength, $F^{\mu\nu},
\tilde{F}_{\mu\nu}$ -- tensor of the electromagnetic field
strength and its dual tensor respectively, $\frac{1}{4M}$  -- the
coupling constant.

The axion rest mass is assumed to be small due to cosmological and
astrophysical constraints  $10^{-6} \, {\rm eV} \leq m_{\rm a} \leq
10^{-2} \, {\rm eV}$, the coupling constant is small either:
$M>10^{10}$ GeV  \cite{sik3}, \cite{khlop}.

The  conversion probability in transverse  magnetic field $B$ is
\cite{sik1} (we use Planck system of units where energy has
inverse length dimensionality in eqns. (\ref{probability}),
(\ref{q}), (\ref{q2})):

\begin{equation}
\label{probability}
P_{\gamma}(L)=2\left(\frac{B}{2M}\right)^2\left[\frac{1-\cos{qL}}{q^2}\right],
\end{equation}
where $q$ is the axion-photon momentum difference:

\begin{equation}
\label{impulse} q=\frac{|m_{\gamma}^2-m_{\rm a}^2|}{2E_{\rm a}},
\end{equation}

$m_{\gamma}$ is the plasma mass of a photon:

\begin{equation}
\label{plasma} m_{\gamma}=0.37\sqrt{n/10^{8} \rm{cm}^{-3}} \, {\rm \mu
eV},
\end{equation}
$m_{\rm a}, E_{\rm a}$ -- the rest mass and the energy of the axion
respectively.

We estimate the probability of the conversion of DM axions to
photons in magnetosphere of a NS using several simplifying
assumptions: 1) velocity of a NS relative to axions is
perpendicular to its rotational axis (dispersion of  axions
velocities can be neglected due to its small value \cite{sik2}) 2)
$r$-dependence for the magnetic field of a NS can be described by
the following relationship $B(r)=B_0(r_0^3/r^3)$, where $B_0$ --
magnetic field strength on the surface of a NS, $r_0$ -- radius
of a NS.

Flux from the axion-photon conversion changes with a period equal
to a half of the spin period of a NS because the dipole axis is
perpendicular to the axion flow twice during one revolution.

If the  plasma mass of  photon is equal to zero the conversion is
severely suppressed. However, density of charged particles in a NS
magnetosphere is quite high and this makes the conversion
possible. For our estimates we use the Goldreich-Julian density.
There are claims that the plasma density in the region of closed
field lines of highly magnetized stars -- magnetars -- can exceed
the G-J density by several orders of magnitude \cite{reaetal}.
This is related to the fact that magnetars have hard tails in
their spectra, discovered thanks to observations aboard Integral
satellite (\cite{gotzetal} and references therein). On the other
hand, X-ray Dim Isolated Neutrons Stars - also known as "The
Magnificent seven", see below -- which we discuss in this paper as
prominent candidates do not have such hard tails, so  we suppose
that one can securely set the value of the plasma density in the
case of the Magnificent seven equal to the G-J density.
 \cite{GJ}:

\begin{equation}
\label{density} n_{\rm GJ}=7\cdot 10^{-2}\frac{B}{T}\,   {\rm cm^{-3}}
\end{equation}

$T$ -- spin period of a NS (in seconds), $B$ -- magnetic field
(Gauss).

$$
n(r)=\alpha_1 B(r)T^{-1}=\alpha_1 B_0 r_0^3 r^{-3} T^{-1},
$$

$\alpha_1=7\cdot 10^{-2} \, {\rm s\cdot cm^{-3}\cdot G^{-1}}$

So, the photon plasma mass depends on radius as follows:

$$
m_{\gamma}(r)=\alpha_2\sqrt{n(r)}=\alpha_1^{1/2}\alpha_2B_0^{1/2}r_0^{3/2}r^{-3/2}T^{-1/2}.
$$
Here $\alpha_2=3.7\cdot10^{-11} \, {\rm cm^{3/2} eV}$. The
conversion takes place when the photon plasma mass coincides with
the rest mass of the axion, $m_{\rm a}=m_{\gamma}.$

The critical radius $r_{\rm c}$ and the critical magnetic field
$B_{\rm c}(r_{\rm c})$  can be derived from the following condition:
$$m_{\gamma}=m_{\rm a},$$
and therefore:
\begin{equation}
\label{critical_r} r_{\rm
c}=\alpha_1^{1/3}\alpha_2^{2/3}B_0^{1/3}T^{-1/3}m_{\rm
a}^{-2/3}r_0
\end{equation}

\begin{equation}
\label{critical_B} B_{\rm c}=\alpha_1^{-1}\alpha_2^{-2}m_{\rm
a}^2T
\end{equation}
For cold axions ($E_{\rm a}\approx m_{\rm a}$, \cite{sik2}) the certain
relation holds:

\begin{equation}
\label{cold}
q=\frac{|m_{\gamma}^2-m_{\rm a}^2|}{2E_{\rm a}}\approx
|m_{\gamma}-m_{\rm a}|\equiv \Delta m
\end{equation}

The difference between the photon plasma mass and the axion mass
depends on the length of axion's path $L$ near the critical point
(the conversion radius) before conversion takes place:

$$
\Delta m \approx
\left|\frac{dm_{\gamma}(r_{\rm c})}{dr}L\right|=\frac{3m_{\gamma}(r_{\rm
c})}{2r_{\rm c}}L
$$

On the other hand, that length $L$ can be determined from the
condition of maximum probability (\ref{probability}):
\begin{equation}
\label{q} qL=\pi\end{equation}

As a result we have:

\begin{equation}
\label{q2} q^2=\frac{3\pi}{2}\frac{m_{\rm a}}{r_{\rm c}}
\end{equation}

After rewriting the expression for the probability of conversion
$P_{\gamma}=B_{\rm c}^2M^{-2}q^{-2}$ with (\ref{critical_r}),
(\ref{critical_B}), (\ref{q}), we obtain:

\begin{equation}
\label{P}
P_{\gamma}\approx20~\rm{G^{-2}}\rm{cm^{-1}}\rm{eV^{3}}\frac{2}{3\pi}\alpha_1^{-5/3}\alpha_2^{-10/3}B_0^{1/3}T^{5/3}r_0m_a^{7/3}M^{-2},
\end{equation}
where coefficient $20~\rm{G^{-2}}\rm{cm^{-1}}\rm{eV^{3}}$ is
needed to adjust eq. (\ref{probability}) written in Planck system
of units to system that is used in conclusive calculations. The
full flux of photons also depends linearly on the critical radius
$r_{\rm c}$ because the amount of axions that propagate through
the region of active conversion increases linearly with growth of
the critical radius. So, finally we can write down our estimate
for the amount of energy that comes from the conversion in one
second, $\dot E$:

\begin{equation}
\label{energy} \dot E\sim \alpha_2^{-8/3}\alpha_1^{-4/3}B_0^{2/3}T
^{4/3}r_0^2m_{\rm a}^{5/3}M^{-2}
\end{equation}

 It is clear that the probability rises sharply with growth of the NS spin
period and the rest mass of axion. However, there are limitations on
$B_{\rm c}$\footnote{Soft Gamma Repeaters (SGRs) are
neutron stars with the strongest known magnetic fields,
$B\sim10^{14\div 15}$ G. Fields of the most
magnetized pulsars do not exceed few$\times 10^{13}$ G.
There is no evidence of
existence of NS with fields stronger then $10^{15}$ G (although the physical
limit of field strength is about $10^{18}$~G).}
and  $r_{\rm c}$ (it can not be less than radius of NS ), so we should specify our candidates for observations for future estimates.

\section{Possibility of observation}
For subsequent estimates we assume the range  of axion rest masses
$0.1 ~\mu\,  {\rm  eV} <m_{\rm a}<10 ~\mu\,  {\rm  eV}$.

Using  (\ref{density}) and  (\ref{plasma}) we can obtain the
following relation:

\begin{equation}
\label{B_crit} B_{\rm c}=10^{10}T\left(\frac{m_{\rm a}}{1\mu eV
}\right)^2~\rm{G}\end{equation}

After  substitution  of typical values of axion rest mass and NS
parameters into the equation for conversion probability (\ref{P})
it can be seen that relevant conversion takes place only for
magnetic fields $B_{\rm c}>10^{11}$ G -- so we need a NS with strong
magnetic field. Also it is favorable to have a NS situated close
to the Solar system with a spin period as large as possible.

In our opinion, the best candidates to produce observable signal due to
axions conversion are X-ray dim isolated NSs.  Seven
objects of this type are known (they are dubbed the Magnificent
Seven (M7), and we use this term below), see a review on isolated NSs in
 \cite{popov}. They possess
very strong magnetic fields (up to $10^{14}$ G) and they are
located not far from the Solar system ($\sim$ 300 pc\cite{poss}
\footnote{The closest object RX J1856.5-3754 is located at $\sim 170$
pc\cite{kerk}}). Their present evolutionary state is not known, yet:
we do not know whether the M7 sources
are similar to normal pulsars or whether
they have already passed this evolutionary stage, and so do not produce
radio pulsar-like emission. We  use the
Goldreich-Julian density of charged particles in magnetosphere for
our estimates. If the density is significantly higher than
the Goldreich-Julian value (about such possibility see, for example,
\cite{lyutikov}), the conversion would be
greatly  depressed, see eq.~(\ref{P}) because it would take place in
area with low magnetic field strength and therefore the
probability of conversion would be tiny.

So, let the magnetic field on the surface of NS be  equal to
$10^{14}$ G, spin period is equal to 10 s and the distance from the
Earth is equal to 300 pc.

For $m_{\rm a}=5  \,  {\rm  \mu eV}$:

$$B_{\rm c}\approx 2.5\cdot 10^{12}    \,  {\rm G}$$

$$r_{\rm c}=3.4 \, r_0$$

$$q^2=\frac{3\pi m_{\rm a}}{2r_{\rm c}}=1.3\cdot10^{-16} \,  {\rm  eV^2}$$

$$P_{\gamma}\approx0.2$$
It is essential to estimate total mass of axions  that fly through
the zone of active conversion ($r<r_{\rm c}$) per unit time to
obtain estimates for the energy of electromagnetic waves from the
conversion \cite{nuss}.

\begin{equation}
 \label{mass} \dot{m}=2\pi (r_{\rm c}-r_0) GM_{\rm NS} \rho v^{-1},
\end{equation}
where $\rho$ --axion density, $v$ -- NS velocity  relative to DM,
$M_{\rm NS}$ -- NS mass. Density of the DM,  $\rho$ (and therefore
the density of axions) is set equal to $10^{-24} \,  {\rm g\,
cm^{-3}} $ \cite{sik2} (the density may be smaller  as such large
values can appear only if a source is in a caustic, but as the
flux depends on it linearly, it can be easily recalculated with
any value of the axion density), velocity was set equal to $v=100
\, {\rm km \, s^{-1}}$ and we use  mass $M_{\rm NS}=1.4 M_{\odot}$
and radius $r_0=10\, {\rm km}$. Total mass of axions propagating
through the "active" region per second is:

$$\dot{m}=2.8\cdot10^2\,
{\rm g\, s^{-1}}$$

The energy that comes from the conversion every second can be
estimated as:

$$\dot{E}=P_{\gamma}\dot{m}c^2=5.4\cdot10^{22}\,
{\rm erg\, s^{-1}}$$

The electromagnetic flux at $300\, {\rm pc}$ from a source might
be $5\cdot10^{-21}\, {\rm erg\, cm^{-2}\,  s^{-1}}$. Radio waves would
have frequencies near central frequency corresponding to the rest
mass of axion $m_{\rm a}$, $f_0=1200$ MHz in the bandwidth $\delta
f=f_0 q/m_{\rm a} =2.8$~MHz. The density of that flux might
be equal to 0.2 mJy.

\begin{figure}[h]
 \includegraphics[]{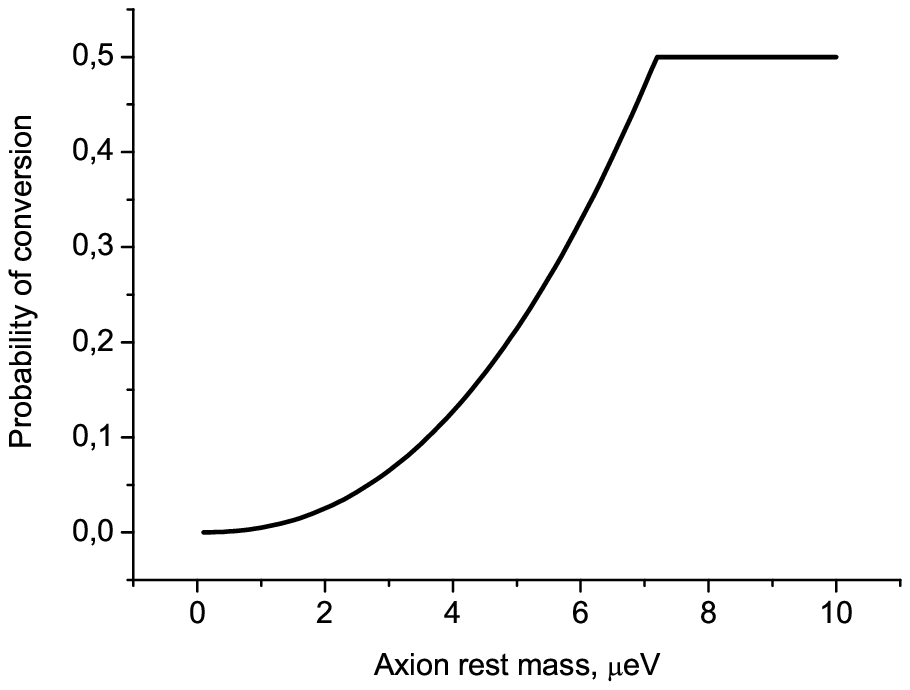}
 \caption{Probability of conversion vs. the axion rest mass.
The probability, $P$,  increases with the mass increase. There is
a saturation at $P=0.5$ because of the process of reverse
axion-photon conversion that effectively suppresses further growth
of $P$. Here and in the figures thereinafter, the magnetic field
on the surface of NS  is  equal to $10^{14}$ G, spin period is
equal to 10 s and the distance from the Earth is set to 300 pc. }
  \label{fig1}
 \end{figure}

\begin{figure}[ht]
 \includegraphics[]{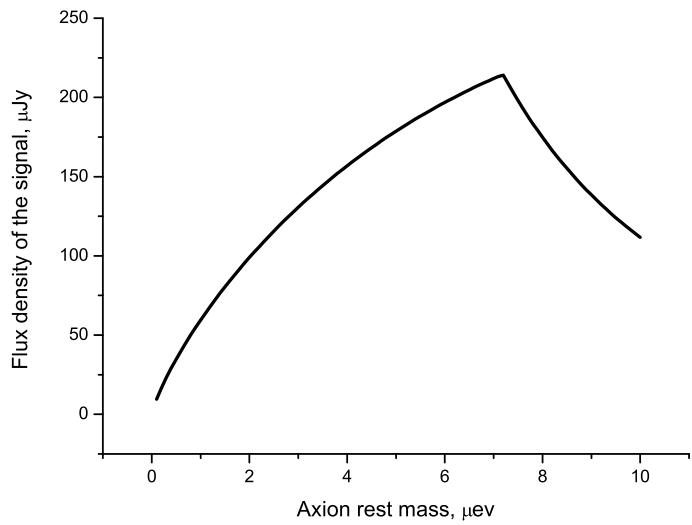}
 \caption{Flux density  of the signal that comes from the conversion of axions of
a certain rest mass.}  \label{fig2}
 \end{figure}

\begin{figure}[ht]
 \includegraphics[]{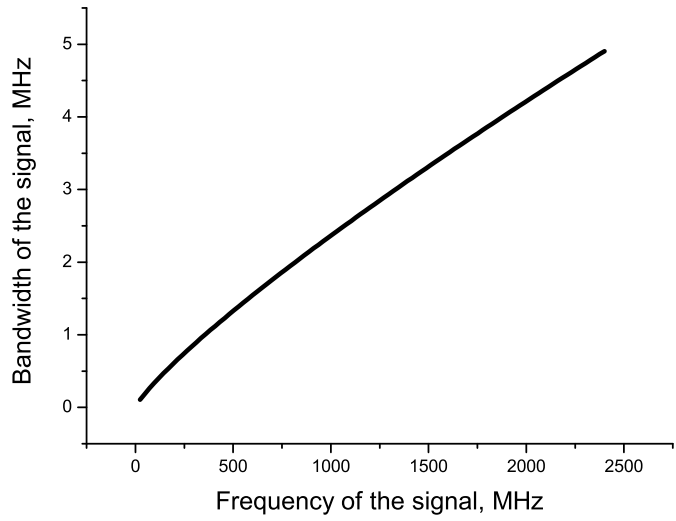}
 \caption{Bandwidth of the possible signal from the conversion vs. the frequency of observation (axion rest mass).
The bandwidth is defined by the magnitude of axion-photon momentum
difference $q(f)$ from the eqs. ((\ref{cold}),(\ref{q})), $q^2\sim
f^{5/6} $}
  \label{fig3}
 \end{figure}

We made estimates for the  values of axion rest mass between 0.1
$\mu\,  {\rm  eV} $ and 10 $\mu\,  {\rm  eV}$. Probability of the
conversion rises swiftly with increase of axion rest mass and
reach its saturation value $P=0.5$ at $m_{\rm a}\approx7 \mu\rm{eV}$
(fig.\ref{fig1}). The predicted observable flux has a sharp peak
at that value of axion rest mass and then steeply decreases (fig.
\ref{fig2}). The bandwidth of the signal from the conversion
smoothly increases with increasing axion rest mass (fig.
\ref{fig3})

The estimates above are only  upper  limits. We do not take into account the
 dipole structure of the magnetic field:
\begin{equation}
 \label{magndipole} B(\mathbf{n},r)=\frac{3\mathbf{n}(\mathbf{n}\mathbf{m})-\mathbf{m}}{r^3},
  \end{equation}
where $\mathbf{n}$ -- the unit vector  along the radius vector
$\mathbf{r}$, $\mathbf{m}$ -- the vector of the magnetic dipole.
Accurate estimates with an exact configuration of the magnetic
field, can slightly reduce our predictions for the flux. Also, the
flux might be smaller because of the reverse conversion, but these
questions require a detailed study of individual cases,  and so we
do not discuss them in this note.

It is necessary to mention the possibility of photons absorption
due to their propagation in high-density plasma near a NS surface.
Indeed, the EM waves of a certain frequency  can not propagate
through regions where plasma frequency exceeds the EM wave frequency.
In the case of a NS that moves through an axion flow away from the
Earth,  converted photons will be reflected  by the magnetosphere
so the sought signal will be weakened. In the opposite case, dense
plasma will act as a mirror, sending converted photons to the
Earth, thus boosting the signal (it might be confined in a solid
angle less than $4\pi$).

Probability of conversion could be further decreased by the effect
of vacuum polarization \cite{gnedin2}. However, the strength of
magnetic field in a region of active conversion is well below the
critical value $5\cdot10^{13}~\rm{G}$. The variability of the
signal depends on the angles $\Psi$ and $\Phi$ between the spin
and magnetic axis and between the spin axis and the velocity
vector of a NS.

\begin{figure}[h]
 \includegraphics[]{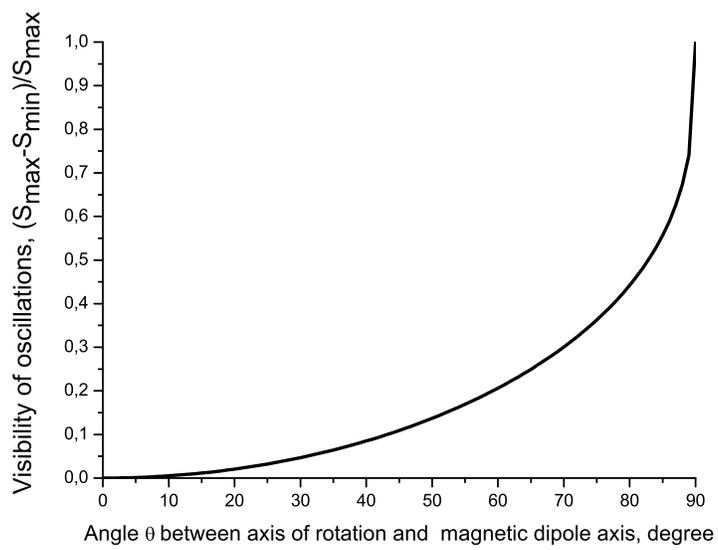}
 \caption{Dependence of the variability of the flux on the angle $\theta$ between the axis of rotation and
the magnetic dipole axis.}
  \label{fig4}
 \end{figure}

In general, the flux variability with time is very complicated and
 should be studied numerically. The figure \ref{fig4} represents
the dependence of the amplitude of the flux variability
$A=(S_{\rm max}-S_{\rm min})/S_{\rm max}$ on the angle $\Psi$ when the
relative velocity of NS to axions is perpendicular to its axis of
rotation.

It is easy to see that the variability will be significant for the
angles close  to $90^{\circ}$ ($P>0.5$ if $\Psi>80^{\circ}$) and
usually it will not exceed 0.1. The period of variation might be twice
 shorter than the spin period of the NS.

At the moment the most stringent upper limits on the radio
emission from the M7 are given by \cite{kond}. At the frequency
820 MHz they are about 10 mJy. At lower frequencies the situation
is less clear. The group from Pushchino \cite{malofeev} announced
detection of pulsed emission from two NSs belonging to the M7. On
the other hand, recent observations with GMRT (Joshi et al.,
poster at COSPAR-2008 and work in progress) do not confirm it. In
the near future LOFAR observations \cite{lofar} can be used to put
better limits or to detect the signal due to axion-photon
conversion. The signal from the conversion will be strongly
depolarized because direction of magnetic field is different in
various parts of active conversion region.
\section{Conclusions} We
suggest to use radio observations of close-by X-ray dim cooling
isolated NSs to search for observational appearances of DM axions.
Whether the axion-photon conversion in NS's magnetic field takes
place, the radio spectrum of the object might have a very distinct
feature -- a narrow spike at a frequency corresponding to the rest
mass of axion. If the coupling constant of the photon-axion
interaction is $M=10^{10}$ GeV, the density of DM axions is
$\rho=10^{-24} \, {\rm g \, cm^{-3}}$ and a NS with $B\sim
10^{14}$~G is located at the distance of 300 pc from the Solar
system,  then the flux density of  signal for axions with rest
mass of $5 \, {\rm \mu eV}$ is as large as several tenths of mJy
at the frequencies $\approx 1200$ MHz in the bandwidth $\approx 3$
MHz.

Acknowledgements. M.P. wishes to thank  Dr. Mikhail Sazhin for
useful comments. S.P. thanks Alexander Gvozdev and Igor Ognev for
discussions. The work of S.P. was supported by INTAS and by
the RFBR grant 07-02-00961.

\end{document}